\documentclass[twocolumn,showpacs,preprintnumbers,amsmath,amssymb]{revtex4}

\usepackage{graphicx}
\usepackage{dcolumn}
\usepackage{bm}
\usepackage{pstricks}
\usepackage{subfigure}

\begin{document}

\newcommand{\ket}[1]{| #1 \rangle}
\newcommand{\bra}[1]{\langle #1 |}
\newcommand{\braket}[2]{\langle #1 | #2 \rangle}
\newcommand{\proj}[1]{| #1\rangle\!\langle #1 |}
\preprint{}

\title{Quantum Error Correction in Globally Controlled Arrays} 

\author{Adel  Bririd}%
 \email{ab373@cam.ac.uk}
\affiliation{%
Microelectronics Research Center, Cavendish Laboratory, University
of Cambridge CB3 0HE, United Kingdom
}%
\author{Simon Charles Benjamin}%
 \email{s.benjamin@qubit.org}
\affiliation{%
Centre for Quantum Computation, Clarendon Laboratory, University
of Oxford OX1 3PU, United Kingdom
}%
\author{Alastair Kay}%
 \email{alastair.kay@qubit.org}
\affiliation{%
Centre for Quantum Computation, DAMTP, Centre for Mathematical Sciences,
University of Cambridge CB3 0WA, United Kingdom
}%

\date{August 4, 2004}

\begin{abstract}
An interesting concept in quantum computation is that of global
control (GC), where there is no need to manipulate qubits individually.
It is known that one can implement a universal
set of quantum gates on a one-dimensional array {\em purely} via
signals that target the entire structure indiscriminately. But
large-scale quantum computation requires more than this: one must be able to
perform efficient error correction, which has
requirements in terms of noise level, time, space
(scaling) and in particular parallelism. Keeping in mind these
requirements, we prove GC can support error-correction, by
implementing two simple codes. We discuss the issues involved in extending our approach to full fault-tolerant computation with this type of architecture.
\end{abstract}
\maketitle

Typical proposals for a solid-state quantum computation (QC)
demand a precise control of both the individual constituent qubits and the interactions between them. The Kane scheme~\cite{Kane} is an archetypal example: in all such approaches one is required to
switch individual interactions `on' and `off'. An alternative is
the idea of \textit{global control} \cite{lloyd, Controlunit,
SimonB, SimonB.2, SimonB.2003}. In this approach, it is not
necessary to individually address qubits. Instead one applies
global signals, e.g. laser pulses, to the entire structure
indiscriminately. With a suitable arrangement of qubits within the
array, it is then possible to discover a sequence of such signals
which have a net effect only at the desired points (i.e. only
affecting specific qubits).

From an experimental standpoint, there are obvious advantages to
such an approach. It can simplify the device structure, removing
the need to have control elements (e.g. metallic electrodes)
directed to each and every qubit. Moreover, in many otherwise
promising QIP candidates, such as molecular scale structures, is
it simply impossible to `plumb in' qubit specific control elements.
The global control paradigm may be the only option in such cases.
Furthermore, even in rare cases where there is no technical
obstacle to fabricating multiple control elements, there remains
the issue that each such element is a potential decoherence
source, typically `dangerously' close to the qubits.

 Lloyd's original GC model
\cite{lloyd, Controlunit} involved a one-dimensional array of
cells, each being a two-state system coupled to its nearest
neighbors with an Ising type interaction. Lloyd employed a regular
pattern of three cell types, $ABCABC...$, where each type has a
distinct transition energy. He demonstrated that one can perform a
universal set of gates with this type of structure, noting that
only the cell on the end need be independently controlled.
Subsequently, Benjamin \cite{SimonB} demonstrated that the same
type of result can be achieved with only two types of cells,
$ABABAB..$, and without the need to distinguish between the
neighbors of a cell. More recently Benjamin has extended the
approach to systems with Heisenberg interactions, initially
assuming that interaction strengths can be collectively switched
\cite{SimonB.2}, and subsequently dispensing with this condition
\cite{SimonB.2003,SimonPreP}.

 In all these variants, there are costs associated with adopting
 the global control principle. Most obviously, they each employ
 an encoding that associates several physical (pseudo-)spins with
 each logical qubit. There is also a corresponding need for each
 logical gate operation to be rendered into several global pulses.
 The severity of these space/time costs increases as
 we consider systems with lower complexity (e.g. fewer cell types)
 or systems over which we have less control (e.g. no ability to
 switch interactions, even collectively). For one extreme case,
 namely a one-dimensional array of two alternating cell types with
 `always-on' Ising interactions \cite{SimonB}, the best known
 encoding requires at least eight physical spins for each logical
 qubit. However, a more profound cost is the loss of parallelism,
 as we discuss later.

The criteria for successful quantum error correction (QEC) \cite{Shor, Steane3} and fault-tolerant
computation \cite{Steane2} have been established in considerable detail. Steane \cite{Steane1} and others
 have shown that besides the basic per qubit error rate, factors like time (number of
steps), space (e.g. the number of qubits used) and in particular
the level of parallelism (number of logic gates performed
simultaneously in the same computational step) must be considered
as well. It is known that one-dimensional structures with nearest
neighbor interactions can support QEC \cite{Ben-Or2}, but the
issue of parallelism is crucial. Aharonov and Ben-Or \cite{Ben-Or}
have employed an elegant analysis to prove that, in the presence
of noise (error rate per quantum logic gate) QC requires a degree
of parallelism which is better than logarithmic in device size.
Their argument involved excluding classes of quantum system than
{\em can} be simulated efficiently on a classical device, and
which, therefore, are not interesting as quantum computers. So, it
is a fundamental requirement that this level of parallelism should
be exceeded in order to perform new quantum algorithms which are
actually infeasible classically. Here we argue that, within
certain reasonable assumptions, globally controlled architectures
can implement error correction with a degree of parallelism that
scales {\em linearly} with system size -- thus exceeding their
condition.

\begin{figure}[!h]
  \begin{center}
    \leavevmode
\resizebox{7.8 cm}{!}{\includegraphics{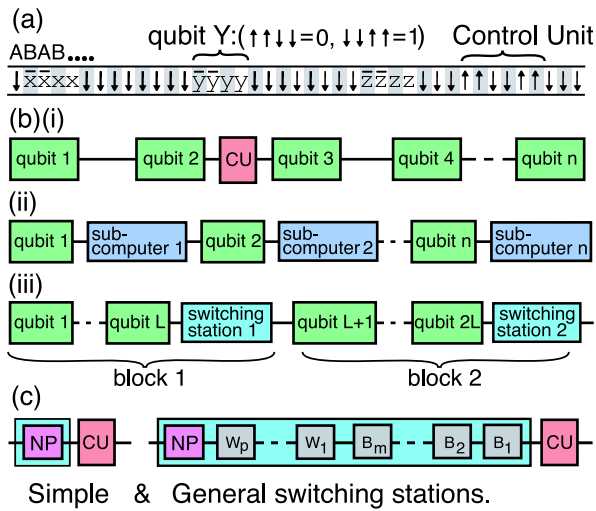}}
\end{center}
\caption{(a) The encoding of qubits and the control unit,
following the scheme in Ref. \cite{SimonB}. (b)(i) Serial model
with a single control unit (b)(ii) Parallel model using
`subcomputer', enabling access to individual qubits (b)(iii)
Model optimized for error correction, with one `switching station'
(SS) for each of the $M$ blocks. (c) The general SS includes $m={\rm
Log}_2(M)$ label bits, a few `working' bits, and a CU.} \label{figure:cellaut}
\end{figure}

Our approach is suitable for all the schemes of Refs \cite{lloyd,
Controlunit, SimonB, SimonB.2, SimonB.2003,SimonPreP}. When we wish
explicitly to count the numbers of steps, etc, we will employ the
scheme in Ref. \cite{SimonB} (depicted in Fig.
\ref{figure:cellaut}(a)). All the schemes share a common tactic to
localize quantum operations to one point (and hence one qubit)
within the device {\em despite} the constraint that all control
signals are sent indiscriminately to the entire structure. They
employ a `control unit' (CU) - not a physical device but rather a
local pattern of states which (in the simplest schemes) occurs in
only one place along the device. With an
appropriate choice of representation for the qubits and for the
CU, it is possible to find sequences of updates to perform a
`toolbox' of basic functions:
\begin{enumerate}
\item \label{sequence1}
A sequence whose net effect is to move the CU with respect to the
qubits, without disrupting those qubits. Then we can position our
unique CU pattern anywhere within the set of qubits.
\item \label{sequence2}
A sequence which has the net effect of transforming the qubit
nearest to the CU, but no net effect on any other qubit. This
allows single qubit gates to be implemented.
\item \label{sequence3}
A third sequence, the least trivial to derive, which implements a
gate operation as the CU moves back and forth between {\em two} qubits.
\end{enumerate}
For initial state preparation and eventual readout, we may assume
that the cell on one end of the device is independently
controllable and readable (perhaps by being physically coupled to
additional devices). Alternatively, we might perform readout by
exploiting a dissipative decay, as mentioned later in connection
with qubit erasure.

It is important to stress that the CU is a
\textit{classical} set of definite 0's and 1's (except when it is
actively involved in performing a quantum gate, as described
above). Therefore, one can employ a relatively crude form of error
prevention for the CU - for example \cite{SimonB} we can cause the state of CU to collapse (effectively, to measure it) when we wish. This is important since an error in the CU pattern
itself could be catastrophic to the scheme - much as a bit flip in
program memory of a conventional computer could cause a crash.  If a CU bit acquires a small element of superposition, due to a slightly imperfect EM pulse for example, a subsequent projection onto the 0/1 basis will recover the correct state with high probability. This can be thought of as a low level ``Zeno'' type of error correction \cite{zenoEffect}. We can illustrate the idea with the following observation: in the scheme of Ref.\cite{SimonB}, at times other than when a gate is actively being applied, it should never be the case that a `lone' $\ket{1}$, i.e. a $\ket{1}$ with both neighbors as $\ket{0}$, exists {\em anywhere} on the device. Then we are free to frequently send pulses that have the effect, ``If you are in state $\ket{1}$ and your neighbors are both $\ket{0}$, then dissipatively reset to $\ket{0}$''. Such measures can stabilize against bit-flips; since the CU is a classical state, it is an eigenstate of the Z operator and is thus already tolerant of phase-flips.

With this simple approach our device has become purely sequential,
since updates have a net effect only at the location of the CU
(Fig. 1(b)(i)). The obvious way to try and recover some degree of
parallelism is to have numerous CU's present at different
locations. Since a parallel algorithm will involve varying
distributions of simultaneous gates, we must somehow
create/destroy CUs. Doing so directly by external intervention
would require local control, so we must instead find a way to
effectively disable a CU by processes internal to the device.
Benjamin \cite{SimonB} proposed an elaborate scheme for doing so,
which indeed recovers a high degree of parallelism: for a device
containing $N$ basic qubits one introduces $N$ CUs, and $N$
uniquely labelled regions called `sub-computers' to enable/disable
those CUs (Fig. 1(b)(ii)). This interesting idea suffers from the
high number of additional cells required. Each qubit would need
$\lceil {\rm log}_2(N)\rceil$ label bits along with a number, $R$,
of auxiliary bits. Taking account of the spacing, which doubles
the number of cells needed, we have $2*4*(\lceil {\rm
log}_2(N)\rceil+R)+10$ cells needed for each subcomputer+CU+qubit.
A device containing 80,000 cells, which in the simple architecture
could support 10,000 qubits, can now support fewer than 900
qubits.

\begin{figure}[!t]
  \begin{center}
    \leavevmode
\resizebox{7.8 cm}{!}{\includegraphics{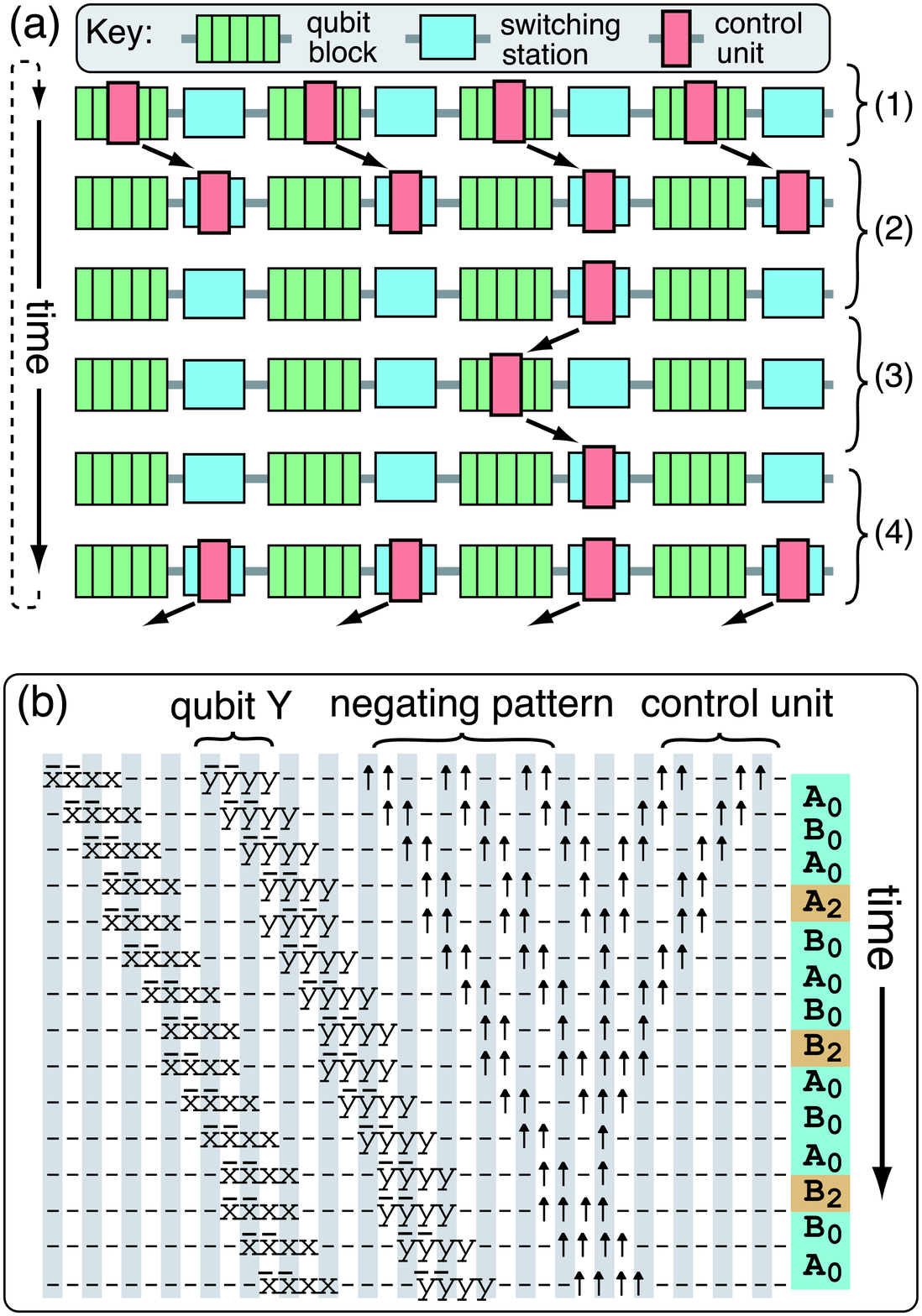}}
\end{center}
\caption{  Upper: Schematic showing the process of alternating
between the error correction phase (1) and the algorithmic phase
(3). This involves switching `off' (2) the majority of the CUs,
and subsequently switching them back `on' (4). Lower: One possible means
of effectively `deactivating' a CU. Here the SS is simply a binary pattern
occupying 10 cells, denoted ``negating pattern" in this figure
(here dashes denote cells in state $\ket{\downarrow}$). In
response to the update sequence shown on the right, the CU is
`absorbed' into the pattern. Note that if all the updates $A_2$
and $B_2$ were omitted, then the CU would pass
\textit{transparently} through. We have designed this process for
the scheme introduced in Ref.\cite{SimonB}, but procedures for the
other GC schemes can be similarly obtained.}
\label{figure:process}
\end{figure}

\begin{figure}[!t]
  \begin{center}
    \leavevmode
\resizebox{8.5 cm}{!}{\includegraphics{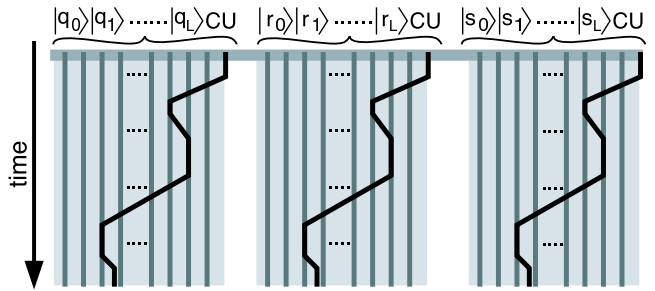}}
\end{center}
\caption{  Schematic of the block encoding process. In response to global pulses, each Control
Unit (black lines) simultaneously performs the
operations of the QEC within a multi-qubit block that encodes a given logical qubit ($q$, $r$, $s$).
The white areas
correspond to the qubits left for spacing and the Switching
Stations. } \label{figure:process2}
\end{figure}

We propose a less costly process, related to this `sub-computer'
idea but optimized to produce a form of parallelism that is useful
for QEC. We describe the process in detail for simple
(non-concatenated) codes and then discuss an extension to fault
tolerant scenarios. The basic concept is to exploit the fact that
typical codes (such as the Steane [[7,1,3]] code, or the Shor
[[9,1,3]] code) involve representing each ``computational qubit"
by encoding it into a block of $L$ adjacent basic qubits. Suppose
that our device contains $N$ basic qubits (each possibly requiring
several cells for its representation), and these basic qubits are
used to represent a smaller number $M$ of encoded qubits.
For simplicity of exposition, we assume that each
encoded qubit corresponds to a distinct block, although this is
not a requirement. During the error correction process, we will
have $M$ CUs active in the entire device (Fig. 1(b)(iii)), one for
each block~\cite{simToLloyd}. Note that the size of a block itself
remains constant once the QEC code has been chosen, the number of
adjacent qubits used to encode each computational qubit being the
same for all of them. The number of computational qubits then
scales linearly with the size of the array. More importantly, the
number of pulses needed to perform any particular operation during
the QEC process is independent of the actual array size, as the
movement of the CU stays confined within each block of constant
size (depicted in Fig. \ref{figure:process2}).

In this way, as we will discuss, we can correct all $M$ encoded
qubits in constant time, independent of $M$. We then `switch off'
the majority of these CUs -- in the simplest scheme, all but one.
This reduced subset of CUs is used to perform step(s) of the
actual quantum algorithm (a Shor factorization, say), before
reactivating all $M$ CUs for another error correction cycle (Fig.
2(a)). The process of activating and deactivating the CUs takes
place in regions which we will call `switching stations' (SS) to
differentiate from the more costly `sub-computers' in Benjamin's
earlier proposal.

It may be helpful to note that during the periods when we are
moving CUs around (eg the first three pulses in Fig. 2(b)), each
pulse moves the CU though a distance of one unit - we are
discretely `clocking' the motion. (In fact, because Fig. 2 employs
the particular scheme of Ref\cite{SimonB}, the qubits also
necessarily move one unit in the opposite direction, so their
separation is reduced by two units per pulse, but for clarity
we simply speak of the CUs moving to their targets). The discrete,
rather than ballistic, motion of these of these entities allows us
to `choreograph' the motion of complex processes, involving
multiple qubits approaching multiple targets, without needing to
be concerned about different entities meeting at slightly
different times.

Consider first the simple case where we wish to switch between $M$
CUs in the device, and just one CU. This can be done with
relatively little cost, either in time or cell count. Each SS
occupies a small number of cells at the right side (say) of each
$L$ qubit block, as shown in Fig. 2(a). All but one of the SS are
identical: each contains a pattern of states that has the property
that it will absorb, or deactivate, a CU when suitable global
pulses are applied (Fig 2(b)). Relative to the CU, this pattern stays in the
SS area and then can be viewed as static, like the qubits. The
process must be reversible so that a CU is emitted, or
reactivated, when the inverse sequence is applied. The
remaining SS is exceptional: it does not act in this way and, in
fact, it can simply be an empty region of the array. Thus when each
of the $M$ CUs passes through the corresponding SS it will be
deactivated - except for the CU passing though the `empty' SS.

The details of exactly how the SS causes a deactivation of a CU
will depend on the exact scheme. For completeness, we briefly
describe some options relevant to the particular scheme of Ref.
\cite{SimonB}. The simple method mentioned above is to literally remove the CU from
the device by using the `negating pattern' shown in Fig.
\ref{figure:process}. Alternatively one could delay the CU, causing it to fail to
`arrive' at the target qubit \cite{SimonB}. A third possibility, perhaps the
most conceptually elegant, would exploit the fact that the scheme
can easily implement gates with multiple control qubits (such as
Control-Control-NOT for example). One would employ a single bit in
the SS and use this bit as the control qubit in a conditional
gate. Thus, an unconditional single qubit gate $G$ in the logical
algorithm would become a Control-$G$ with the SS as the
controlling bit, and similarly a Control-NOT would be implemented
as a Control-Control-NOT, etc.

Having removed all but one of the CUs, we can now proceed to
perform any manipulation we wish with the single remaining CU. In
fact we will perform the next step of the overall quantum
algorithm (a Shor factorization task, we have supposed). It will
only be possible to perform a certain number of steps before it
becomes necessary to apply another error correction cycle: then we
simply apply the reverse of the deactivation process to recover
all $M$ CUs. Ideally we would wish to have at least enough time
between error correction cycles to perform an arbitrary
two-qubit gate between remote qubits. Because we are assuming a 1D
array with nearest neighbor interaction, the time required for
this scales with $N$ - thus for a very large device one might not
be able to complete such a gate. In this case one could resort to
performing a series of swaps, after
successive EC cycles, to bring the qubits closer.\\

 Within each
error correction phase, the $M$ active CUs are each associated
with one of the $M$ encoding blocks (Fig. 2(a) step 1). They
simultaneously perform the same series of qubit operations within
each block to implement the EC code (Fig. \ref{figure:process2}).
Recalling that
each block is also of constant size (for simple EC), and that $M$ is linearly proportional to the device size, we conclude that we need only apply a fixed number of pulses
to accomplish the error correction over the entire device. That is, we achieve the ideal of linearly scaling parallelism which of course exceeds
the condition on parallelism mentioned above.

Our global pulses necessarily cause exactly the same behavior
within each block, whereas the errors (if any) that occur in each
block will vary. However, this can be dealt with by internalizing
the process of applying the `fix' within the EC algorithm. Let us
consider the quantum state $|Q_P\rangle$ of an encoded block,
subject to decoherence from the environment. Such decoherence can always~\cite{Nielsen} be
characterized by the X (bit-flip) or Z (phase-flip) types of
error, or a combination of the two (Y$\equiv$ZX). In the following lines we
neglect normalization for the sake of clarity. Note that a
`collapse' to a specific state of the environment can occur at any
time without altering the result. The general process of
interaction with the environment takes the initial state
$|Q_P\rangle|E_0\rangle$ to
$$
|Q_P\rangle|E_0\rangle+\delta_X|Q_X\rangle|E_1\rangle+\delta_Y|Q_Y\rangle|E_2\rangle+\delta_Z|Q_Z\rangle|E_3\rangle
$$
At the beginning of the error correction process we introduce
ancilla qubits, initially all in state $\ket{0}$, onto which the
error syndrome is placed. This results in a particular auxiliary
state for each of the different error types:
\begin{eqnarray}
&\ket{Q_P}\ket{E_0}\ket{A_0}+\ \ \ \ \ \ \ \ \ \ \ \ \ \ \ \ \ \ \ \ \ \ \ \ \ \ \ \ \ \ \ \ \ \ \ \ \ \ \ \ \ \ \ \ \ \ \ \ \ \ \ \ \ \ &\nonumber\\
&\delta_X\ket{Q_X}\ket{E_1}\ket{A_1}+\delta_Y\ket{Q_Y}\ket{E_2}\ket{A_2}+\delta_Z\ket{Q_Z}\ket{E_3}\ket{A_3}&\nonumber
\end{eqnarray}
The circuits we are using here apply a specific correction
conditional on the error syndrome given by the ancilla qubits. The
original state is then recovered for each type of error syndrome:
$$
\ket{Q_P}\left(\ket{E_0}\ket{A_0}+\delta_X\ket{E_1}\ket{A_1}+\delta_Y\ket{E_2}\ket{A_2}+\delta_Z\ket{E_3}\ket{A_3}\right)
$$
Subsequently the ancilla qubits are erased back to $\ket{0}$, making them available for another syndrome extraction. There is thus no need for a syndrome
`measurement' in the commonly described sense
\cite{Nielsen}.

\begin{figure*}[t]
\centering \resizebox{17.5cm}{!}{\includegraphics{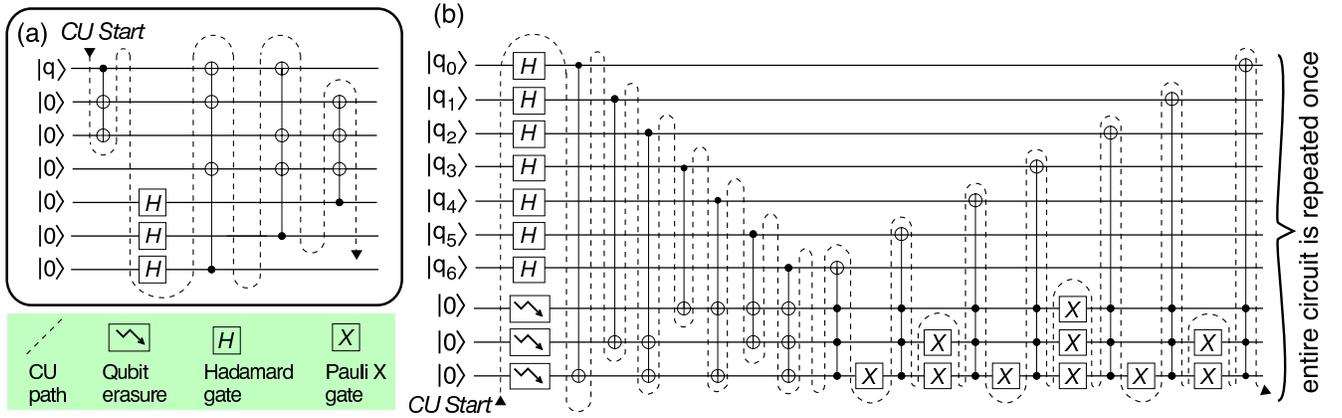}}
\caption{ Encoding (a) and EC circuits (b) for the Steane [[7,1,3]]
code. The computational qubits of the circuits are in the same
order, from top to bottom, as they are on the cell chain from right
to left (c.f. Figs. 1 \& 2). EC involves syndrome extraction and
subsequent correction for both phase-flip and bit-flip errors. The
movement of the CU back-and-forth through the array is indicated by
the dashed line. } \label{figure:steane2}
\end{figure*}

\begin{figure*}[t]
\centering \resizebox{17.cm}{!}{\includegraphics{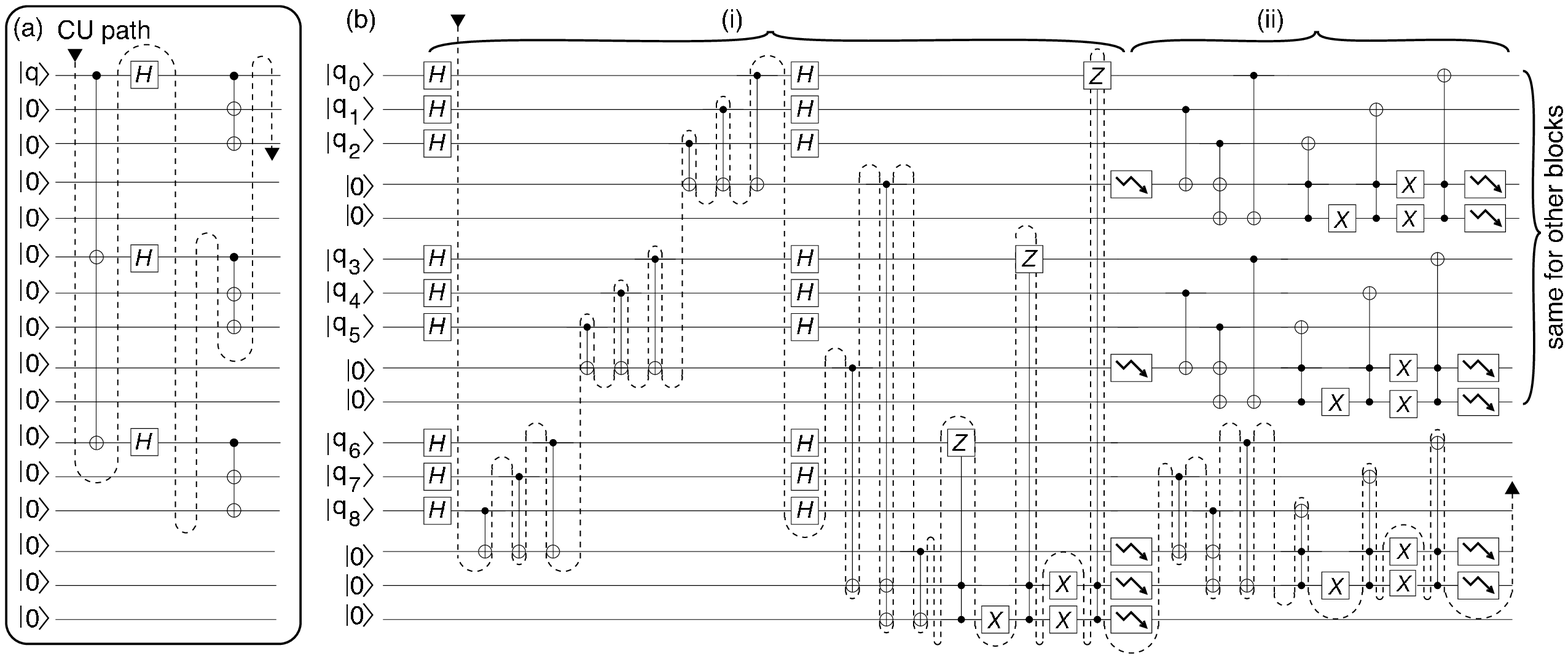}}
\caption{Encoding (a) and EC circuits (b) for the Shor [[9,1,3]]
code.  Phase-flip errors are resolved in section (i), and bit-flip
errors in (ii). The notation is as defined in the key in Fig.
\ref{figure:steane2}.} \label{figure:shor2}
\end{figure*}

The behavior of the CU within a given block is shown explicitly in
Figs. \ref{figure:steane2} and \ref{figure:shor2} for the Steane [[7,1,3]] code and the Shor [[9,1,3]]. These simple (non-fault
tolerant) QECCs each correct a single error. The movement of the
CU is indicated in order to show the importance of finding an
efficient route. To formally minimize the path of the CU, one
would need to specify the complexity of the various types of gate,
which depends on the particular physical model
\cite{lloyd,Controlunit,SimonB, SimonB.2, SimonB.2003}.

Due to its
movement through the block, the CU could be seen as potentially
dangerous for the scheme if it carries any error, as this could be
propagated to all the qubits. Here we should stress again the
importance of the CU being a \textit{classical} pattern of states,
which makes it much more resilient to errors in that we can {\em
use dissipation} to stabilize it. Thus we would aim to achieve a
degree of stability in the CU more akin to that of conventional
bits rather than qubits. Additionally, for the particular encoding
scheme we have adopted here \cite{SimonB}, there is a strong sense in which the
CU moves \textit{transparently} through the intervening qubits:
this transparency is not `constructed' by performing swap
operations, for example, but occurs `naturally' as the CU and
qubit collide under the simple driving sequence (pulses $A_0, B_0,
A_0, B_0 ...$). There is minimal interaction between the CU and
any qubits passed, and consequently the risk of error propagation
via the CU is also minimized.

In these simple circuits, and in more complex fault-tolerant
procedures, it is crucially important to be able to reset (erase)
qubits. Here we assume that this can be achieved by a mechanism
analogous to the one proposed in Ref. \cite{SimonB} for efficient
measurement. We envisage that at least one of the cell `types' has
an unstable third state $\ket{\rightsquigarrow}$, which rapidly
decays to the ground state. Then we can initiate an erasure via a
special form of one-qubit gate operation (symbolized by an arrow
in the circuits, Fig. \ref{figure:steane2} and \ref{figure:shor2})
using:

\centerline{$\mathbf{U}=
\begin {pmatrix} 0&0&1\\
\noalign{\medskip} 0&1&0\\
\noalign{\medskip} 1&0&0\\
\end {pmatrix}$ in the basis \{$\downarrow$, $\uparrow$,
$\rightsquigarrow$\}.}

The use of such auxiliary state(s) could potentially introduce
problems of `leakage' out of the computational subspace. In order to negate this risk,
 the third state must be chosen to be in a very
distinct energy level, separated from the qubit states by an
energy gap that prevents any spontaneous jump of the quantum
states to this level. This state can then only be reached by a
precise manipulation via the quantum logic gate U. As a physical example, one might consider the
qubit states as the eigenstates of an electron spin in a quantum dot (or molecule)
while the transient state is an optically excited state (e.g. an exciton) that can only
be reached from one of the spin states due to Pauli blocking~\cite{pazy}.
Such a state would decay extremely rapidly, and could not be ``acidentally'' excited
by ambient thermal excitations, even at room temperature.

A physical process such as this will allow erasure
of the ancilla qubits after the error syndrome has been
successfully used to correct a bit/phase error (see for example
rightmost of Fig. \ref{figure:shor2} (i)). Once reset to the
ground state, they can be used again within a given EC process,
thus reducing the total number of auxiliary qubits required per
block. Figures \ref{figure:steane2} and \ref{figure:shor2} both include optimizations of this kind.

In Table 1 we make a comparison of the
efficiency of the two codes illustrated in Figs. \ref{figure:steane2} and  \ref{figure:shor2}. However we should emphasize that the scheme we describe here is completely independent of the size of the
chosen EC code, and then perfectly scalable to more complex or
simpler ones. Indeed it would be highly desirable from an
experimental perspective to start with a much simpler model of
code (simply correcting bit-flip errors for example) to demonstrate the
feasibility of implementing the basic scheme.
For the explicit pulse count, we assume the physical model in Ref. \cite{SimonB}, since this is
the `worst case' in the number of cells per qubit. To get an idea
of the time scale of a correcting process, the table gives a
calculation of the number of pulses needed to perform all the
quantum operations of each circuit, within one encoding block and
with one CU. Erasure can be considered here as a particular
one-qubit gate operation.
\begin{table}
\caption{\label{tab:table1}Characteristic figures for the number of primitive global pulses
needed for error-correction, for each of
the two codes implemented.\\}
\begin{ruledtabular}
\begin{tabular}{lcr}
Correction step&Steane
[[7,1,3]]&Shor [[9,1,3]]\\
\hline
Encoding & 464 & 397 \\
Syndrome measur./recovery& 3622 & 2955 \\
Total& 4086& 3352 \\
\end{tabular}
\end{ruledtabular}
\end{table}

With the inclusion of the ancilla qubits, the Steane code
is encoded on blocks of 10 qubits (requiring 80
cells), and the Shor code on blocks of 16 qubits (128
cells). A simple
sketch of the global movement of the CU through the array is
described in each circuit, designed to minimize
its movements between the different gates. We must
stress that the circuit design and the movement of the CU are not aggressively optimized
for minimum pulse count - there are improvements that can `shave off' a few pulses at the cost
that the circuits become more cumbersome to illustrate. Therefore the figures in the Table should
be regarded as characteristic rather than definitive.
In Appendix II we make some remarks to illustrate the
counting process that we employed.
The Table shows a modest speedup of about 20 percent in favor of
the Shor code, which is balanced by the fact this code needs more
qubits per block to be implemented.

\begin{figure}[t]
  \begin{center}
    \leavevmode
\resizebox{8 cm}{!}{\includegraphics{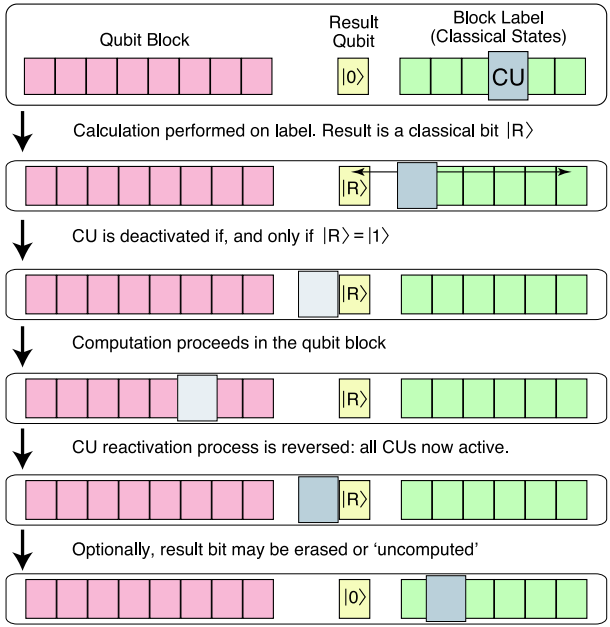}}
\end{center}
\caption{ This is a schematic representation of a single block of qubits and their switching station, incorporating the idea of a classical label of states to identify the switching station, so that an algorithm can be run to selectively enable/disable a subset of the CUs. }
\label{figure:labels}
\end{figure}

In the scheme described above, we switch between $M$ CUs for the
error correcting phase, and just one CU for the algorithm phase
(Fig. 2(a)). This allowed us to use trivially simple switching
stations, that unconditionally deactivate, or reactivate, a CU
(Figs. 1(b)(iii) \& 2(b)). Consequently each SS can be a fixed
small size, and they require a constant proportion of the entire
device as it scales. In this respect the scheme is efficient: it
does not require significant additional spatial resources versus
the basic global control scheme. However, in another sense it is
inefficient: during the algorithm phase we have only a single CU
at our disposal, and therefore our potentially parallel device is
acting as a purely serial computer. We can reach a better
compromise between spatial and temporal efficiency by
introducing a specialized form of the `labels' concept mentioned in Ref.
\cite{SimonB}. Now each SS gains a binary label which we can exploit to differentiate
it from neighboring stations (Fig. 1(c)). In the strong limit of using unique labels, this therefore requires Order($M {\rm Log}_2(M)$)
additional cells in total (although shorter, repeating labels can be useful, as we demonstrate in Appendix I). The process of switching from the error
correcting phase to the algorithmic phase is now less trivial, although
the actual error correcting phase is not affected at all by
the labelling. We apply a sequence of global pulses that causes
each of the $M$ CUs to simultaneously move to a SS, and there to
perform a small computation ${\bf C}$. The computation applied in
each SS is of course identical, being driven by the same pulses,
but the data on which the computation acts, i.e. the binary label,
is unique. Therefore the outcome of the computation will vary from
one SS to the next. This outcome is a binary variable represented
by either altering, or not altering, the pattern of states
responsible for (de)activating the CU. Then when the CU
subsequently moves through this region (c.f. Fig 2(b)), it may or
may not be deactivated, depending on ${\bf C}$ and the label of
that SS. In this way we can switch from all $M$ CUs being active
during the error correction phase, to some chosen subset of all
$M$ CUs in the algorithmic phase. This is illustrated in Figure \ref{figure:labels}. Note that the time $\tau$ associated with the label computation must be less than O($N$)
otherwise it would have been quicker to perform the parallel gates
in series. One could not enable/disable a completely arbitrary
sub-set of the $M$ CUs under this time constraint \cite{patExist}, so
our procedure does not efficiently implement a completely general
arrangement of gates. However, there are a great many useful
distributions of CUs which do correspond to fast $\tau$. Two examples are discussed in Appendix I.

We now discuss some of the issues involved in extending the above
ideas to fully fault tolerant (FT) processing. The above arguments
have demonstrated that a high degree of parallelism can be
harnessed for EC; in fact this parallelism can also be optimally
exploited for operating on concatenated codes. In Appendix I we
give some explicit examples of how this can be achieved.
We conclude, therefore, that 1D globally controlled structures can achieve
adequate parallelism for FT processing. However a second issue is that of
performing operations {\em between} encoded blocks in a robust
way. Typically fault--tolerant implementation of such a gate
involves performing a measurement over multiple qubits and
applying an operation depending on the result \cite{Nielsen}. To
formally establish fault tolerance, it is necessary to show how to
`internalise' such operations with an algorithm, in much the same
way as syndrome extraction and correction was performed with an
algorithm. Fortunately Aharonov and Ben-Or have indeed shown
(Theorem 4 of \cite{Ben-Or2}) how to perform these gates for a
subset of CSS codes. Importantly, they use only operations which
are standard elements of computation, and require only the
additional ability of adding and discarding qubits, which is
equivalent to the qubit erasure process that we have described. As
one might expect, operating without the measurement process
increases the overhead in terms of the number of ancillas required
and apparently results in a more severe error threshold (although it is emphasized
that their result is not optimal). Hence, for a suitable choice of
error correcting code, a globally controlled architecture not only
has sufficient parallelism to allow fault tolerant quantum
computation (which in turn allows a quantum algorithm to be run to
arbitrary accuracy), but has known techniques that detail exactly
how to perform such operations.

It is important to stress, however, that the CU must behave very reliably if is not to be an ``Achilles' heel'' in an FT implementation. If a permanent error were to occur on a CU, then this could be catastrophic for the calculation since this would not only affect a single qubit in one level of concatenation, but would also affect every other level. However, as has already been noted, the CUs are classical states and can thus be stabilised against decoherence, e.g. by dissipation, more simply and effectively than quantum states can. We note that a temporary error in a CU would be less of a problem. In the worst case, a single encoded qubit at a given level of concatenation acquires an error. The QECC of the next level of concatenation can correct for this single error (once the error has been removed from the CU). Nevertheless, one must regard an error in the CU as being as potentially destructive as a bit error in the program of a conventional computer. Thus, in physical systems where the CU cannot be sufficiently stabilized (for example, any system in which flip errors are a dominant source of noise) the route to FT processing which we outline here {\em would not be suitable}. It is an interesting question whether some variant of the CU idea can be conceived, within which  one could immediately detect and correct CU errors. This is perhaps a direction for future work.

The authors would like to thank Andrew Steane for useful
conversations. This work was supported by the EPSRC and the DTI
under the foresight LINK project. SCB acknowledges the support of the Royal Society.

\newpage
\newpage
\noindent {\Large \bf Appendix I}
\smallskip

Here we demonstrate the potential use of the `labelled switching station' idea by describing two particularly interesting arrangements of control units, both of which are relevant to fault tolerant computation with so-called {\em concatenated codes}. In such codes the QECC is applied recursively to `encode the codes', possibly through multiple levels. The first arrangement we discuss is the fundamental pattern for such processing, i.e. a pattern of one CU per block, where that block may now consist of a hierarchy of lesser blocks. The second pattern is a local group of CUs suitable for performing efficient `bit wise' operations on encoded blocks, without decoding them.

For the basic, non-fault tolerant EC scenario described in the
body of the paper, we were able to switch between a single CU and
a CU for every block of qubits. Such a block consists of $L$
computational qubits (including the necessary ancillas, but not
including the SS). Now however we wish to be able to choose
whether we have a single CU or a CU every $L^i$ SSs, where $i$
denotes the level of concatenation of error correcting code
($i\geq 0$). This will be possible if we place a (non-unique)
classical label in the region of each SS when initialising the
quantum computer. Let us assume that we intend to use a total of
$p-1$ levels of concatenation of our QECCs. Then the number stored
in the label of the $i^{th}$ SS is given by the prescription
$$
SS_i=\sum_{j=1}^{p-1}R(i \text{ mod } L^j)
$$
where the function $R(x)$ returns 1 if $x=1$ and 0 otherwise. The number $p$ is specially reserved to only be at a single location, which would otherwise have contained a $p-1$ label -- this will facilitate switching to a single CU when necessary. We emphasise that the above expression is not a calculation that is {\em run} on the quantum computer, it simply defines the label arrangement which is placed in regions of the computer during initialization. These states are then exploited during the subsequent computation.

\begin{figure}[!t]
  \begin{center}
    \leavevmode
\resizebox{7.5 cm}{!}{\includegraphics{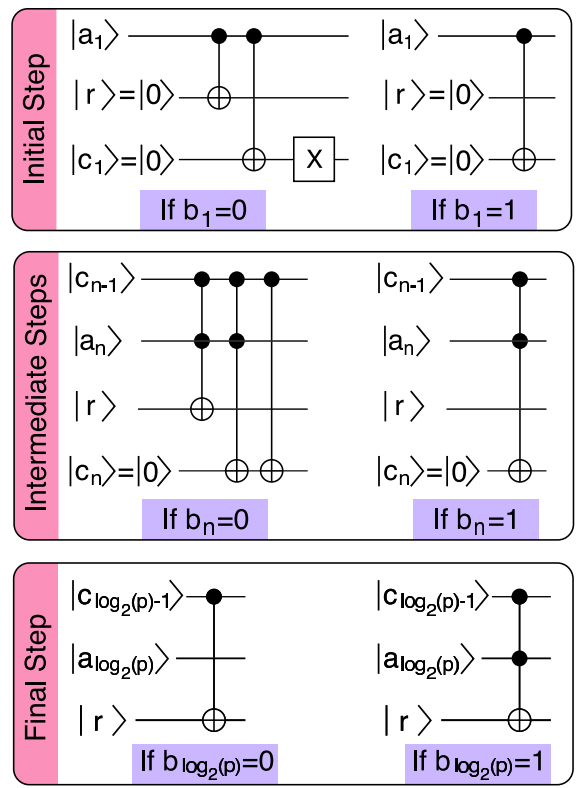}}
\end{center}
\caption{Algorithm to implement a set of CUs suitable for fault tolerant concatenated codes, as described in the text. The symbol $a_x$ represents the $x^{th}$ most significant bit of the number stored in the label. $r$ contains the result of the computation on the label. The bits $c$ are the ancillas that mark whether we have a subsequent recursion, and thus act as a control in the next step.}
\label{figure:fault_calculate}
\end{figure}

Recalling that the algorithm is, at all times, `sent in' by global
pulses, at a given moment during error correction we will be
operating at a certain level of the concatenation, say the $b^{\rm
th}$ level, throughout the device. To switch on a CU every $L^b$
SSs ($0\leq b\leq p-1$), we just run a program that returns 0 if
the value of the label is less than $b$ and 1 otherwise. The
result of this then determines whether or not to `deactivate' a CU
at that SS location. Earlier we discussed the various means by
which a CU can be effectively deactivated. For the present
discussion it is convenient to assume we are employing the third
mechanism we outlined, namely making a single bit in the SS act as
a control bit in subsequent qubit gates. We shall denote the
number stored in the SS label by $a$, and the $x^{th}$ most
significant bit of the number by $a_x$. Fig.
\ref{figure:fault_calculate} shows how to implement a suitable
algorithm using ancillas, $c$ (which we regard as part of the label space). The binary label itself will be stored on
$\log_2(p)$ bits and, given the qubit erasure process, $U$, then
we only require 2 ancillas. We only require $O(\log_2(p))$ steps
in the program to decide which CUs activate/deactivate (given that
the program will run on every SS simultaneously). Recall that $p$ characterizes the total number of levels of encoding and therefore it must be a modest number in any plausible device -- thus our running time of $O(\log_2(p))$ is very fast. The result of
this algorithmic test of the label is the single classical
`outcome' bit in each SS, denoted by $r$ in Fig.
\ref{figure:fault_calculate}. The algorithm shown in that figure
is a general procedure that will work for any $b$. For specific
$b$ it
should be possible to customize the algorithm, making it yet more efficient.

The method presented here can be adapted to different patterns of
CUs, by organising the numbers stored in the SSs differently. The
only requirement is that there exists a hierarchy in the patterns
of CUs such that, to switch between one pattern and another
requires either the activation or deactivation of a subset of CUs,
not both. We shall therefore use the same algorithm for the second
desirable arrangement of control units, which we now discuss.
Suppose we were able to activate and deactivate not just a single
CU in a block, but some form of super-CU, a dense local patterning
of control units capable of performing efficient bitwise
operations on a single encoded qubit, and acting on all the
constituent qubits simultaneously. For example, in the encoding
shown in Fig. \ref{figure:shor2}, this would mean leaving on 9
CUs, in the pattern of 3 on, 2 off, 3 on, 2 off, 3 on (nothing
needs to be done to the ancillas). To generate this pattern at the level of
individual qubits, we
would need to provide a SS for every such qubit, so in fact it
would be inefficient to implement this for a basic QECC. However,
for concatenated codes, we {\em can} make use of the idea without
additional resources.

We aim to create a single super-CU for each level of concatenation {\em above the lowest}. To do this, we set the labels on the switching stations equal to
$$
SS_i=p-\sum_{j=1}^{p-1}\delta(i\leq L^j)\prod_{k=1}^{j-1}R'(i \text{ mod } L^k,k)
$$
where, in the case of the Shor code of figure \ref{figure:shor2},
$L=16$ and $R'(x,k)$ returns $1$ for $1+r\times L^{k-1}\leq x \leq (r+3)\times L^{k-1}$,  where $r\in\{0,5,10\}$.
Note that since we have no super-CU at the lowest level of
concatenation, each operation would need to be repeated 9 times, starting
on the unencoded qubits $1,2,3,6,7,8,11,12,13$ each time. (This is however a small
cost compared to the dramatic device size increase that would be required for a SS for
each qubit.)
The
prescription for $SS_i$ is of the form $p-f(i)$ because we now
want to activate CUs as we go up the levels of concatenation,
instead of deactivating them as we did in the previous example.
Note that when we are computing on an encoded qubit, we also have
to act on encoded ancillas (section 5.4 of \cite{Ben-Or2}), so the
3-2-3-2-3 pattern of the Shor code needs to be carried over to all
levels of concatenation.

We might also like to be able to manifest a {\em repeating}
pattern of these super-CUs across the device, but such a pattern
is not consistent with the requirements for the algorithm of Fig.
\ref{figure:fault_calculate}. That is not to say that such an
algorithm doesn't exist. Any single pattern of CUs for
concatenated codes can always be realised in a SS with $p$ qubits.
The $i^{th}$ qubit simply indicates whether or not that particular
SS should be switched on or off in the $i^{th}$ level of
concatenation. This only takes up a constant proportion of the
device size, and does not require a computation to be performed on
the SS. Whether any encoding is possible should then be obvious
from the `missing' numbers. If all the numbers from 0 to $2^p -1$
are present on different labels, then, obviously, no encoding of
the labels will be able to store the required information more
efficiently.

We cannot perform all operations in a bitwise fashion and so we also need to retain the ability to switch to the control of a single CU every $L^i$ SSs, and to having only a single CU. This is trivial to do, since all we need to do is have two labels (combining to form a single, larger label), using the two systems of label patterning already specified. Depending on what type of state we desire, we choose which half of the pattern to perform the algorithm on.

The discussion in this Appendix has simply highlighted two of the many possible uses of the labelled-SS concept. We hope that these examples suffice to show that the idea is a powerful one.
\smallskip
\smallskip
\smallskip

\noindent {\Large \bf Appendix II}
\smallskip
\smallskip

Here we make a few remarks to illustrate the counting method that
we employed to obtain the pulse totals quoted in the Table.  Recall that we are adopting the
scheme of Ref.~\cite{SimonB} since this is the `worst case' in terms of size costs (a consequence of
the extreme simplicity of the $ABAB$ model). We
distinguish the ``approach pulses", which only move the CU from
being adjacent to one qubit to being adjacent to another, from the ``operation pulses" which modify the state of
the qubit. For example, for a one-qubit logic gate, starting from
the stage where a CU is next to the target qubit a total of 15 pulses are required: 8 pulses
to perform the operation followed by the first 7 {\em in reverse
order} to restore the CU pattern. In the case of a controlled operation between, say, the
first (control) and the fourth (target) qubit, starting from a CU
next to the control qubit, 5 pulses are needed for the CU to interact with the control. Several ``approach pulses" (2*4) follow to bring the
CU next to the target. Then 9 pulses are needed to perform the
operation, and we reapply the first 8 pulses in reverse order as
before. We must then reapply the earlier pulses (approach+ encoding)
in reverse to complete the process of restoring the CU to its original form.
However, it is
efficient  to retain the altered form of the CU if
several qubits are subject to the same control qubit: one can then
move directly to those other targets. The circuits are designed to
employ this optimization wherever possible.


\begin{thebibliography}{99}
\bibitem{simToLloyd}This is in fact reminiscent of an idea presented by Lloyd in his original
work on these structures \cite{Controlunit}, but here we applying it to the specific issue of QEC.
\bibitem{patExist} There must be patterns of CUs which cannot be produced using a sequence of less than
O($M$) updates, given that the number of distinct updates is
limited. Otherwise we would have a means of compressing $M$
arbitrary bits, corresponding to the $M$ enabled/disabled CUs, into
less than O(N) symbols.
\end{thebibliography}
\end{document}